# The Genomic Code: The genome instantiates a generative model of the organism

**Kevin J. Mitchell[1] and Nick Cheney[2]**

[1]Institutes of Genetics and Neuroscience, Trinity College Dublin
Kevin.Mitchell@tcd.ie
[2]Department of Computer Science, University of Vermont
ncheney@uvm.edu

## Abstract (110 words)

How does the genome encode the form of the organism? What is the nature of this *genomic code*? Inspired by recent work in machine learning and neuroscience, we propose that the genome encodes a generative model of the organism. In this scheme, by analogy with variational autoencoders, the genome comprises a connectionist network, embodying a compressed space of "latent variables", with weights that get encoded by the learning algorithm of evolution and decoded through the processes of development. The generative model analogy accounts for the complex, distributed genetic architecture of most traits and the emergent robustness and evolvability of developmental processes, while also offering a conception that lends itself to formalisation.

## Highlights (868 characters including spaces)

- The generative model concept captures the indirect, distributed, and non-linear relationship between information in the genome and the form of the organism.
- The genome embodies a compressed representation in a space of latent variables – the DNA sequence itself, which encodes a connectionist gene-regulatory network
- The latent variables collectively shape an energy landscape that constrains the self-organising processes of development so as to reliably produce a new individual of a certain type.

- This encoding is robust and evolvable, and explains the independent selectability of traits, drawing on the idea of multiplexed disentangled representations observed in artificial and neural systems.

- Finally, it offers a conception that lends itself to formalisation, both of empirical data from systems biology and for simulation of artificial life *in silico*.

**Outstanding Questions (1824 characters, including spaces).**

- What is the "data format" of the generative model? What physical elements and relations comprise the latent variables of the model?

- How does learning by evolution differ from artificial training algorithms such as back-propagation? How is credit assignment achieved?

- How are developmental processes co-adapted alongside genomes different from decoders trained via exemplars?

- Do gene regulatory and signal transduction networks themselves comprise layered structures that can progressively decode the generative model? Or is this progressive decoding realised through the cellular divisions of embryogenesis?

- How does the genome encode developmental trajectories between cellular states, modifying the generative model and constructing new decoders along the way?

- How much of the heavy lifting is done by the self-organising biophysics of the biochemical and cellular materials? In what ways do genomic variables constrain such processes?

- Can this framework help formalise a holistic, systems view of developmental processes? Can a view based on discrete molecular functions and logical operations be reconciled with a dynamical systems view of gradients, fields, and landscapes?

- How important is noise in these schemes? Is it a feature or a bug?

- What kinds of circumstances or pressures drive compression and the emergence of disentangled representations for different traits?

- Can this framework lead to better formal models of epistatic interactions and genotype-phenotype relations for complex traits? Can it help in understanding the evolution of qualitatively novel phenotypes?

- How best can this kind of indirect, distributed encoding be implemented in silico in the creation of artificial life? Will this provide more flexibility, or opportunities for biological inspiration, than current schemes?

**Introduction: The genomic code**

The fundamental phenomenon of genetics is that like begets like. Cats have kittens and dogs have puppies. This is despite, in each case, new individuals starting out as a single, undifferentiated cell – the fertilised egg, or zygote. The complex form of the parents is thus not transmitted to their offspring – instead, it must be reconstructed in each new individual. Clearly then, the zygotes of cats and dogs must contain some substance – the genetic material – that is inherited from the parents, which somehow directs the development of the growing organism so as to produce a new token of the feline or canine type. The question is: how? What is the nature of this genomic encoding of form?

Over the years, various metaphors have been used to conceptualise the nature of what we will call the *genomic code* – i.e., how the genome specifies the form of an organism [1, 2]. These include, among others, a codescript [3], blueprint [4], program [5,6], recipe [7], or resource that the developing organism can draw on [8]. One of the most enduring of these, especially common in popular science treatments, is the idea that the genome constitutes a “blueprint” of the organism (e.g., [4]). This metaphor conveys the idea of a detailed but miniaturised plan that can in some way be referred to, in order to direct the development or construction of a pre-specified final product.

However, the metaphor quickly falls down in several ways [9]. First, an architectural or engineering blueprint is *isomorphic* with the desired product – that is, distinct parts of the blueprint correspond directly and specifically to distinct parts of the product. In this way, the blueprint concept is almost preformationist, with the genome containing a direct mapping of the final product. Second, a blueprint does not usually contain instructions on *how to build* the object in question – it only has information on what it should look like when completed. This clearly leaves a major question unanswered – how the processes of development are specified so as to yield the desired outcome. And finally, a blueprint typically specifies an object in such detail as to be effectively deterministic, leaving little room for the kind of variability in developmental trajectories and outcomes that is typically observed, even in genetically identical organisms raised in highly controlled, effectively identical environments [10].

An alternative metaphor is that of a “program” [5,6]. In this view, the genome does not contain endpoint information in the way that a blueprint does. Instead, it encodes *algorithmic information* – a set of instructions or steps that will reliably lead to the production of some outcome. This analogy was first introduced in 1961, separately, by evolutionary biologist Ernst Mayr [11] and by geneticists Francois Jacob and Jacques Monod [12].

For Mayr, it encapsulated and legitimised the teleological idea of purpose, or end-directedness, as a fundamental principle in biology. A zygote undergoes a series of developmental steps *towards the end of* producing a new individual of the species, and the information specifying those steps must somehow be encoded in the genome. On a mechanistic level, Jacob and Monod drew on their work on gene regulation in bacteria to infer that differential regulation of gene expression might also underlie the specification of the diverse cell types observed in multicellular organisms.

The “program” concept seems to better capture the emergent nature of the relationship between genotype and phenotype. In particular, it encompasses the key idea that whatever potentiality is encoded in the genome, it can only be realised through the processes of development. Those developmental processes can be described as a series of steps, occurring in a stereotyped order, directed towards some end. This corresponds well with the traditional meaning of the word “program” as: “a set of related measures or activities with a particular long-term aim” (Oxford Languages).

However, it is important to distinguish between the observed “developmental program” and the supposed underlying “genetic program” [5]. The former term simply describes the phenomenon that needs explaining – the observable steps of embryogenesis leading to a species-typical outcome. The latter term purports to provide the necessary explanation – the means by which the developmental program is encoded.

But the word “program” also carries some unsupported connotations. Regarding the means by which the steps of development are encoded, “program” is in modern times also defined as: “a series of coded software instructions to control the operation of a computer or other machine” (Oxford Languages). The usage of the term “genetic program" may thus seem to imply a regular, explicit, and interpretable set of instructions, logically laid out and executed in series. As with a blueprint, it can be taken as implying a kind of isomorphism, this time between elements of the genome and elements of the developmental program. Moreover, it again suggests a kind of algorithmic determinism, with all the details somehow spelled out in advance. Neither of these properties is observed.

A variation on the "program" idea is that of a "recipe", as used in baking, for example. Here, you have some set of ingredients – material stuff that must be brought together – as well as a set of instructions for how to do that. These preparatory steps themselves are somewhat like a program, but once the baking starts, the way the system develops is determined by the chemistry and physics at play. This naturally leads to more variability in the precise outcome of development. No matter how detailed your recipe, you can't bake the same cake twice [7]. The outcome is not specified in full detail, nor are the processes themselves. The recipe doesn't need to encode all the relevant chemistry and physics – those are just given. Instead, the conditions are just set in such a way that the processes tend to happen in a robust and reproducible fashion so as to reliably produce an acceptable outcome. For an organism, this would mean landing within a viable phenotypic range.

The "recipe" notion thus avoids some of the deterministic, rigid connotations of a "program". It feels more *organic*. In particular, it appeals more to notions of self-organisation, which offer a different perspective on where the information resides that governs the emergence of the ultimate form of the developing organism. An important aspect of this view is that the genome does not have to encode the specifics of all the developmental processes in a detailed program. It just has to set the conditions in a certain way and let the self-organising nature of the cellular systems do most of the work. The sequence of the genome can constrain the direction of self-organisation, but a lot of the informational heavy lifting is offloaded to the developing system itself.

This jibes with some alternative perspectives, notably from Developmental Systems Theory (DST) and aligned traditions, which view the genome not as a privileged store of information that somehow actively *directs* the development of the organism, but as a "resource" that the developing organism can draw on [8]. Importantly, DST sees the genome as just one resource among many that are inherited, including the cell itself, its environment, and even aspects of culture in some species. Variation in any of these elements can bear on how development proceeds and, in the DST framing, none of them is seen as privileged. This view is more ecumenical and avoids the sense of the genome actively driving development. However, it remains decidedly vague.

Indeed, none of these metaphors provides the means to formalise or operationalise the role of the genome in the encoding of organismal forms. They are too vague to offer any kind of theoretical foundation that could help us model empirically observed developmental processes, including their evolution, or simulate such processes in artificial systems.

Moreover, the current metaphors fail to provide any insights on these crucial questions:

- How does information get encoded in the genome and what is this information *about*?
- What is the “data format” of this information in the genome?
- How does such information get decoded through the processes of development?
- How does genotypic *variation* relate to phenotypic *variation*?
- How does the genetic architecture of various traits arise? (i.e., whether they are affected by many variants or only a few, whether such variants affect specific traits or many, whether they interact in simple or complex ways, and so on).
- How do the underlying encodings evolve through time?

Here we propose a new model inspired by work in artificial intelligence and neuroscience: the genome as a generative model of the organism. This concept seems to more aptly capture the relationships between genotypes and phenotypes, provides at least broad answers to the questions above, and is readily formalisable. (We note that Michael Levin and colleagues and Karl Friston and colleagues have recently independently made similar observations [13-16]).

## The generative model metaphor

The “job description” of the genome is to somehow embody a set of parameters that will lead to the development of a new organism of the right sort from a fertilised egg. In abstract terms, this is similar to the problem solved in machine learning by “autoencoders”. These are neural networks that are trained to produce new instances of text, images, or other data, which are similar, but not necessarily identical, to those from a training distribution. If, for example, the autoencoder is trained on images of horses, then it can output novel images that have the typical characteristics of horses.

This is achieved using a particular architecture of layers of artificial “neurons”. This architecture comprises roughly two halves, firstly an input side – the encoder – which parses input data through multiple layers, with progressively fewer units per layer, until it reaches a central layer with the fewest units. This bottleneck forces a compression of the information in the input data into a lower-dimensional “representation” [17,18]. The output side – the decoder – then reverses this process, decompressing the information in such a way as to generate a new token of the relevant type. The system is trained, in a self-supervised way, to minimise a loss function, which is the distance of some mathematical characterisation of the generated object from the objects in the training distribution (i.e. the quality of the reproduction).

The crucial element in this process is the compression into a low-dimensional representation. This forces the system to learn the abstract statistical regularities that define the type of object in question and to encode these abstract features (e.g., of “horsiness”) in a space with limited information capacity, in a way that most efficiently captures the essence of this object, and which can then be decoded [18,19]. At the same time, the decoder has to learn the right algorithms to decompress this information to generate a new token. In many cases, this leads to an encoding in the compressed layer of “latent variables” that have little direct relationship (i.e., no linear or independent mapping) to the features of the objects on which the autoencoder has been trained. Importantly, the system thus does not merely learn a low-dimensional description of the features of the objects input into this system. Rather, it comes to instantiate a *generative model* – an encoding of latent variables and the means of decoding them to generate new instances of the class of objects seen by the model (Figure 1).

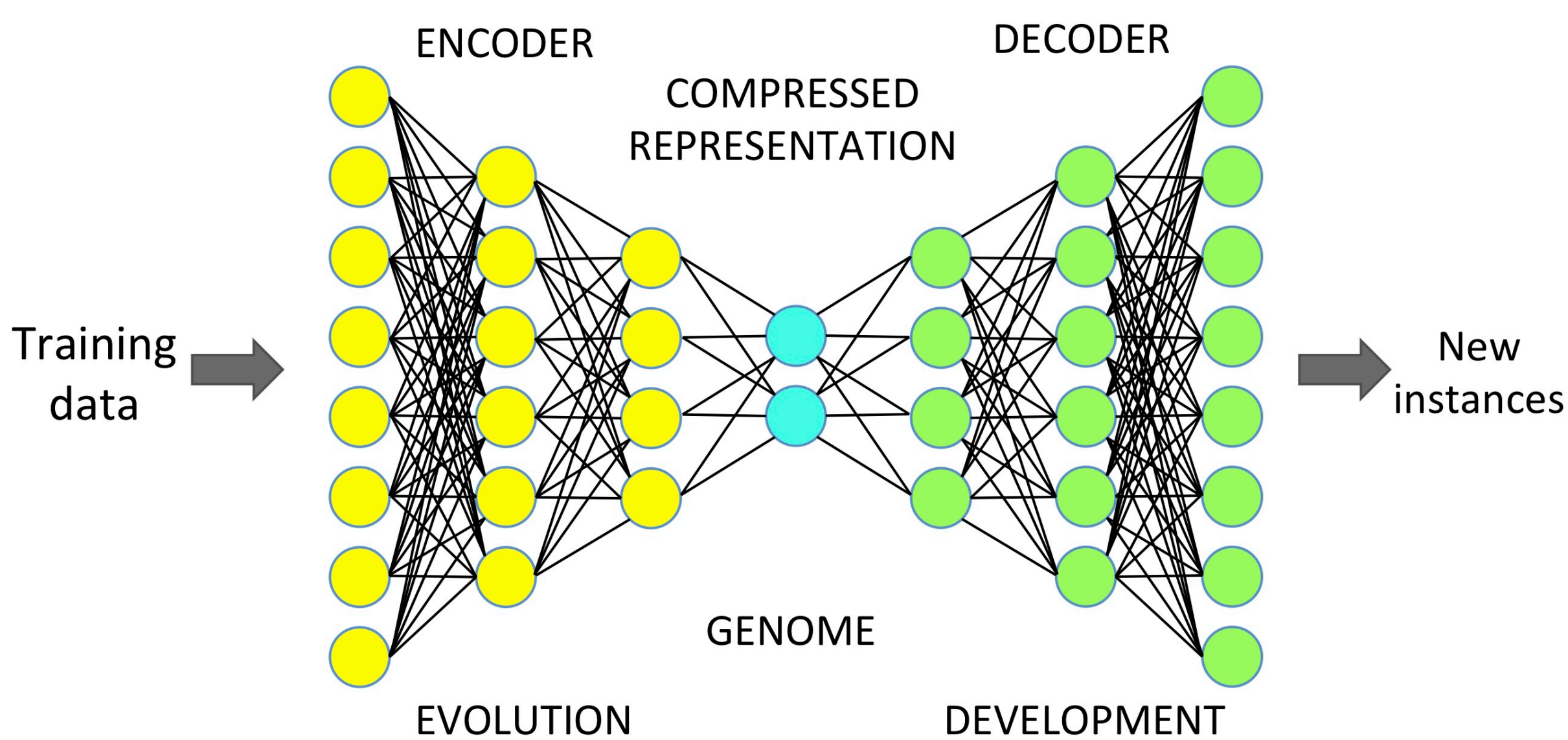

**Figure 1. An analogy with variational autoencoders.** VAEs learn a compressed representation of their training data, leading to a generative model encoded in latent variable space (a “compressed representation”), which can be decoded to produce novel instances of a learned type. By analogy, evolution acts as an encoder, leading to a compressed representation in the genome (a generative model

of the organism), which can be decoded through the processes of development to produce a new individual of a given species.

This has obvious possible parallels to how information about the form of the organism is encoded in the genome. First, the detailed three-dimensional structure of an adult animal cannot simply be replicated. Instead, all that information is massively compressed into just a single cell, with its single copy of the genetic material, from which the pattern is *re-produced*. In this process, the genome comprises an information bottleneck – it does not contain enough information, mathematically speaking, to specify the number and position and type of every cell of the organism [8,20].

The genome does not contain direct endpoint information at all, in fact. In a sense, it contains algorithmic information [21] that governs how developmental processes proceed [22,23]. But even this is not encoded directly or discretely, with distinct parts of the genome specifying instructions for distinct developmental processes. Rather, this encoding is indirect, distributed, and non-linear – the latent variables embodied by the genome *collectively* constrain biochemical interactions such that certain cellular and developmental processes tend to occur in certain ways [9, 24-26].

An additional important element of some autoencoders – called *variational autoencoders* – is that they do not encode latent variables as specific values but rather as probability distributions [27,28]. The decoder is trained to produce a sensible output by sampling from these distributions, thus necessarily producing new tokens that vary idiosyncratically in their details, while still conforming to the abstract type (such as a horse). There is a clear correspondence to natural systems, which must buffer noisy molecular processes or genetic variation so as to still produce viable and potentially reproductively fit offspring [10,29]. This property relates to robustness and evolvability [10,30], as we discuss below.

In the rest of this paper, we explore the elements and implications of this model.

**What are the latent variables in the genome?**

The most obvious thing that is encoded in the genome is the sequence of proteins. The deciphering in the 1960s of “the genetic code” that relates the sequences of DNA bases to the sequences of amino acids in the corresponding proteins was a scientific triumph that set the stage for the rise of molecular biology [31]. This code is linear and direct and means that each gene (i.e., each protein-coding “transcription unit”, in molecular biology terms) is, literally, a kind of blueprint for a specific protein. However, the deciphering of the genetic code for proteins did not by itself reveal *how the genome as a* *whole codes for the form of an organism*. The sequences of the proteins themselves are only part of the picture.

Equally important (or perhaps more so) are the sequences that encode how those proteins are expressed. This “code” is far less regular and discrete. It is not just a linear text that can be directly translated with fixed, universal rules into another linear text – it is not a cipher at all, in fact. It is, rather, part of a dynamic control system with combinatorial, contextual dynamics that allows cells to regulate their own biochemistry under many different conditions. In multicellular organisms, this includes making different cell types in different parts of the organism, thus constraining the processes of development and morphogenesis that shape its ultimate form. There are thus several questions to ask: (i) how is the regulation of expression of specific genes encoded?; (ii) how is this coordinated across the genome in each cell?; and (iii) how do interactions between the resultant cell types ultimately lead to the emergence of species-typical forms?

The genome must encode (or constrain) all these processes and their outcomes, but with only the sequence of DNA nucleotides as the information-bearing elements. DNA is an extraordinarily chemically inert molecule, which is why it is so stable (making it an ideal medium for information storage). It does play a part in chemical reactions as a catalyst, however – specifically as an adsorption catalyst. That is, by bringing together other molecules, it encourages them to interact in ways that would otherwise be highly unlikely. Indeed, the only thing directly encoded in DNA sequences is *differential affinity* – for proteins, RNA molecules, or DNA or RNA nucleotides.

It is this differential affinity that directs regulation of gene expression. It is instantiated in the sequences of the regulatory RNAs and proteins (transcription factors, chromatin proteins, splicing factors, translational regulators, and so on) and the sequences of various regulatory elements they bind to, either in the DNA or in the encoded RNA. Collectively, the affinities of these interactions broadly determine which genes are expressed, to what levels, and in which parts of the developing embryo, thus driving patterning and coordinating cellular differentiation.

The latent variables are thus embodied by the DNA nucleotides themselves, which, over sequences of varying length: (i) encode RNA and protein molecules that do the work in the cellular economy, including the regulation of gene expression; and (ii) comprise binding sites for these regulatory factors. In that sense, the latent variables are inherently *relational*, because they arise from the affinity between trans-acting factors and cis-regulatory elements, which are encoded by separate pieces of DNA. With respect to the form of the organism, these variables are “latent” because the relationship of the genomic sequence to the form of the organism is distributed, non-linear, and extremely indirect.Though just as the generative model in machine learning has many different layers of “latent representations,” each at a different level of abstraction, the organismal form may be represented at many levels of abstraction (e.g. as latent features comprising motifs within the DNA, rather than individual nucleotides). We choose to focus on the analogy between the bottleneck layer of the VAE and DNA nucleotides for the sake of emphasizing storage and compression, rather than an argument that this is the strictly or solely corresponding level of abstraction.

**The decoder: generative models of growth and development**

In the analogy with artificial systems such as variational autoencoders, the cells of the developing embryo play the part of the decoder – the part of the machine learning model which gradually decompresses the latent vector into the full image. Deep neural nets are designed with multiple layers, traditionally arranged in a hierarchical fashion with connectivity existing only between adjacent layers. Progressive processing of data through these layers in the encoder is what enables abstraction of general features and compression into a latent variable space. In the decoder, the sequential progression from the compressed latent space to the resultant image occurs in the opposite direction via sequential layers of higher-and-higher-dimension.

One crucial point of difference between machine learning models that generate new tokens of some type and the processes of multicellular development that generate new individual organisms, is that the latter have to build their own decoders along the way. Not just once, but over and over again. And as they do, they also change the variables in the generative model itself.

To begin with, the chromosomes of the zygote are not just naked DNA, but packaged into chromatin, making some protein binding sites accessible and others inaccessible. This changes the landscape of latent variables. In addition, the zygote inherits a specific complement of nuclear, cytoplasmic, and membrane proteins – the elements that actively do the decoding. The initial conditions of the generative model are thus set by these factors inherited by the cell. (It is important to note, however, that the chromatin modifications on the egg and sperm genomes as well as the complement of cellular proteins in the egg are themselves directed by the genomic sequence in the prior generation).

The same thing happens through cell divisions, with changes to the initial conditions and the configuration of the decoder each time new cell types are generated. This determines which elements of the generative model will be active in any newly generated cell, allowing the decoder to generate the appropriate cell type. The genome thus encodes a multiplicity of models and matching decoders, which are brought into play as embryonic development proceeds. The temporal sequence of cell types in development may thus provide an analogue to structural layers in deep neural nets, progressively processing the data in the latent variable space, with each step determining the nature of the data that are passed to the next one.

The different state spaces that define various cell types are thus encoded in a multiplexed fashion. At least part of the reason they don't interfere with each other is that they are read out by different proteins in different cell types at different times, induced by new signals along the way that are interpreted by new cell states. The program thus builds the new interpreters along the way, as well as adding epigenetic marks to the genome itself. The generative model instantiates directionality to these transitions [32], with each transition revealing latent states that were not reachable before, and, in turn, creating new transition probabilities [26,33,34].

All this cellular differentiation also has to be spatially and temporally coordinated on the scale of the whole organism [35]. This means that the generative model instantiated in the genome must also constrain the developing embryo so as to direct the processes of proliferation, differentiation, cell signaling, and morphogenetic movements that collectively lead to the emergence of the three-dimensional form of the organism [13,14]. These processes rely on all kinds of physical parameters and self-organising dynamics that are not encoded in the genome, and that don't have to be [24,25,35-38]. This starts with the physics driving protein folding, but also includes the biophysics of the cytoskeleton, of adhesive forces, of membrane tension, and so on, which collectively shape the morphology of cells and the morphogenesis of tissues. The genome can take all the relevant physics and chemistry as given. All it needs to encode are sets of constraints that channel these processes along certain trajectories, locally, through cell divisions, and globally, in the emergent morphogenesis of the whole developing organism.

**Gene regulatory networks, attractor states, and energy landscapes**

The ways in which the latent variables in the genomic sequence can constrain and direct development have been modelled and conceptualised in various ways.  First, the collective regulatory interactions encoded in the genome can be depicted as gene regulatory networks – graphs that capture the positive and negative interactions between various regulatory factors and from regulatory factors to their effector targets [39,40]. The regulatory elements of any given gene can be thought of as performing logical operations (AND, OR, NOT, etc.) on their “inputs” – i.e., the pattern of transcriptional regulators present and active in the cell [41,42]. The activity of these regulators is often sensitive to some kind of signal or state. The classic example is the *lac* operon in E. coli, which is transcribed only when glucose is absent (such that the CAP transcriptional activator is active and bound to the DNA) AND lactose is present (such that the lac repressor is inactivated and not bound to the DNA) [11,12].

However, the *lac* operon is unusual in that the genes comprising it are either not transcribed at all or transcribed at very high levels – i.e., their activation state is effectively binary: ON or OFF. Most gene regulation is more graded, from LOW to HIGH (not digital, but analog). In addition, transcription is inherently probabilistic and fluctuating – expression levels do not reflect a smooth rate of mRNA production but rather the probability of the gene being transcribed at any moment [29,43]. Genes thus do not act like clean, isolated Boolean operators but rather as continuous, noisy, and dynamic elements in a connectionist network [41,44,45].

Once such a network gets large enough it becomes too complicated to model as a set of interconnected logic gates that determine the static state of the system. Indeed, in development, specific proteins are “re-used” in different parts of the embryo in combinatorial and highly context-dependent ways, to the extent that it becomes impossible to associate specific factors with specific cell types or tissues in any kind of causally isolated way. Instead, the entire genomic network can be thought of as an interlocking dynamical system that can tend toward a variety of possible stable attractor states, but that also will tend to follow stereotyped trajectories through state space over time, given certain conditions (Figure 2) [26,41,45-48]. A dynamical model also captures the fact that the regulatory interactions at play do not all occur instantaneously. They play out through time, often at different rates.

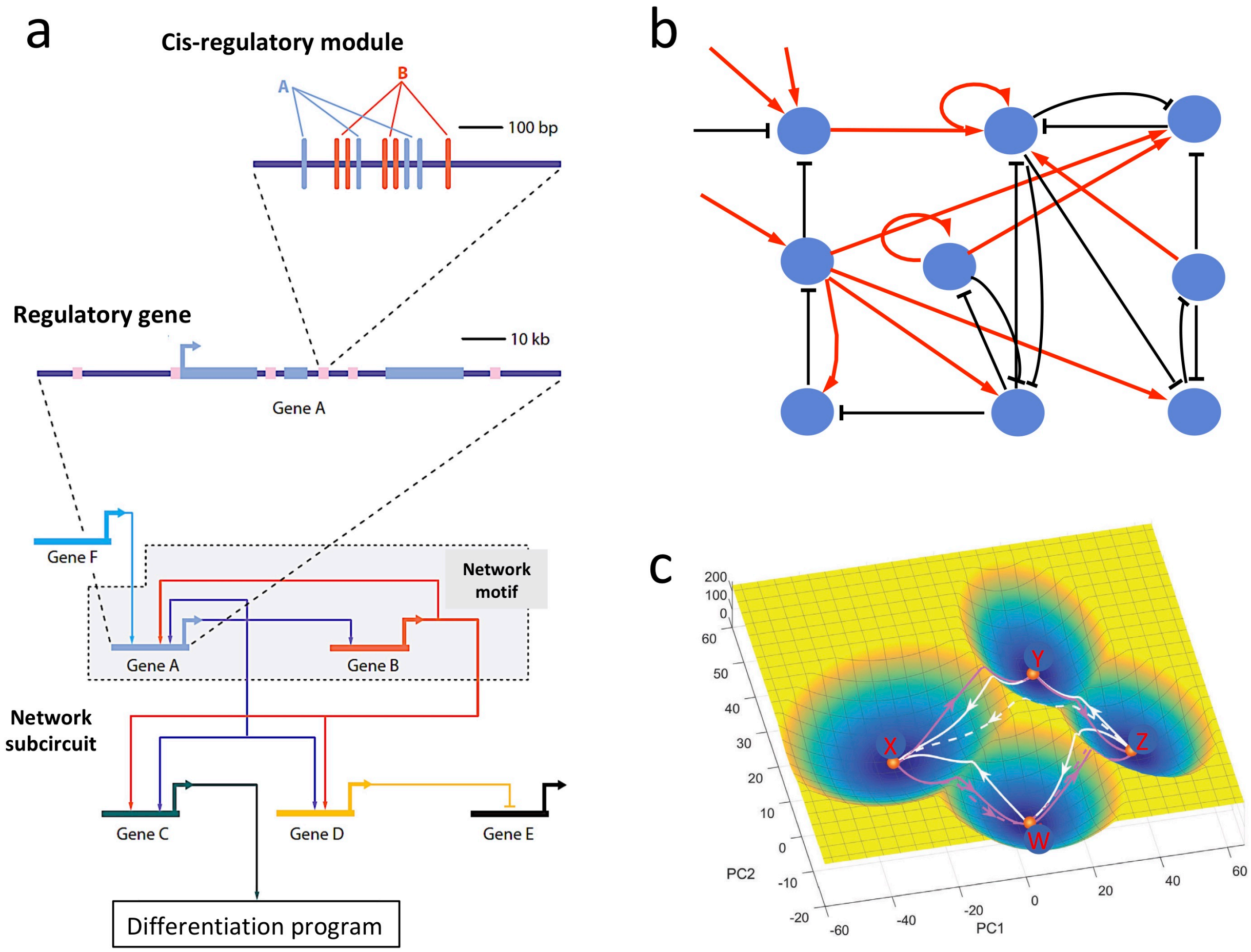

**Figure 2. Gene regulatory networks.** Panel (a) shows the hierarchical arrangement of gene regulation, starting with cis-regulatory DNA elements of a gene (Gene A), comprising binding sites for Gene A's own product, Protein A, as well as the product of Gene B, Protein B. Most genes contain many such regulatory elements. Gene A encodes a regulatory protein, which regulates its own transcription as well as several other targets. The gene regulatory network can be depicted as a series of motifs and subcircuits carrying out interconnected logical operations, ultimately leading to some program of cellular differentiation. Panel (b) shows an alternate view of a gene regulatory network, depicted as a weighted graph or connectionist network. Panel (c) shows how such a network embodies a dynamical system that can generate a landscape of attractor states. (Panel (a) reproduced with permission from [39]. Panels b and c modified from [48]).

In artificial systems, these kinds of complex interactions are modelled as generating an "energy landscape" – a multi-dimensional contoured surface, with peaks and valleys, with different points on this landscape representing possible states of the system [49,50]. At any time, the system will tend towards states with the lowest "energy", with the configuration of the landscape funnelling the system away from unstable peaks and towards one or other of the stable valleys (i.e., the attractor states). The benefit of this depiction is that it incorporates progress of the system through time – not just that various states are possible, but that the system will tend to pass from one such state to another along constrained trajectories.

Conrad Waddington published just such a visualisation – his famous "epigenetic landscape" – in 1957, to describe how cellular differentiation works [51]. He depicted the cell as a ball, rolling down the landscape as development proceeds, being channelled into one of several possible valleys, representing different cell types. One important feature of this scheme is that it is probabilistic. The depth of the various valleys represented the *likelihood* of a given cell adopting a given fate, but at the entrances to these valleys, when the landscape was still pretty flat, small amounts of noise might shift the ball one way or the other. The actual outcome for any individual cell might thus not be fully determined, though the statistical outcome over many such cells would be set by the relative probabilities.

While the picture of the ball rolling down this landscape is well known, the picture of what is going on beneath the landscape is less so. In Waddington's scheme, the landscape is like a huge sheet of canvas, which is pulled down in certain positions by a system of pegs and guy ropes. Each peg represents a gene, and their collective influence – with many genes pulling on the canvas at any one position and individual genes pulling on multiple points – is what generates the shape of the landscape (Figure 3).

The pegs and ropes underlying Waddington's landscape would comprise the latent variables of the model. Crucially, their relationship to the output is combinatorial, non-linear, non-isomorphic, and indirect – and further complexified by the positive and negative feedback loops abundant throughout development. In addition, the system is not forced into particular states – it is merely configured in a way that constrains its self-organising processes. In a sense, the genes define where the system *can't* go.

This visual metaphor is useful as a kind of cartoon of how we can think of what is happening during cellular differentiation and development. But the analogy with generative models suggests that it may actually be formally correct. Generative models mathematically create just such energy landscapes, with the decoding algorithms constraining the outputs to positions in feature space with the lowest energy (and thus the lowest "reconstruction error" in the case of autoencoders) (Figure 3) [49,50]. Importantly, these landscapes are typically smooth and continuous mathematical surfaces, such that objects near each other in feature space are most similar, with features varying smoothly across the landscape. This has implications for how traits may smoothly evolve by altering the shape of the decoding landscape.

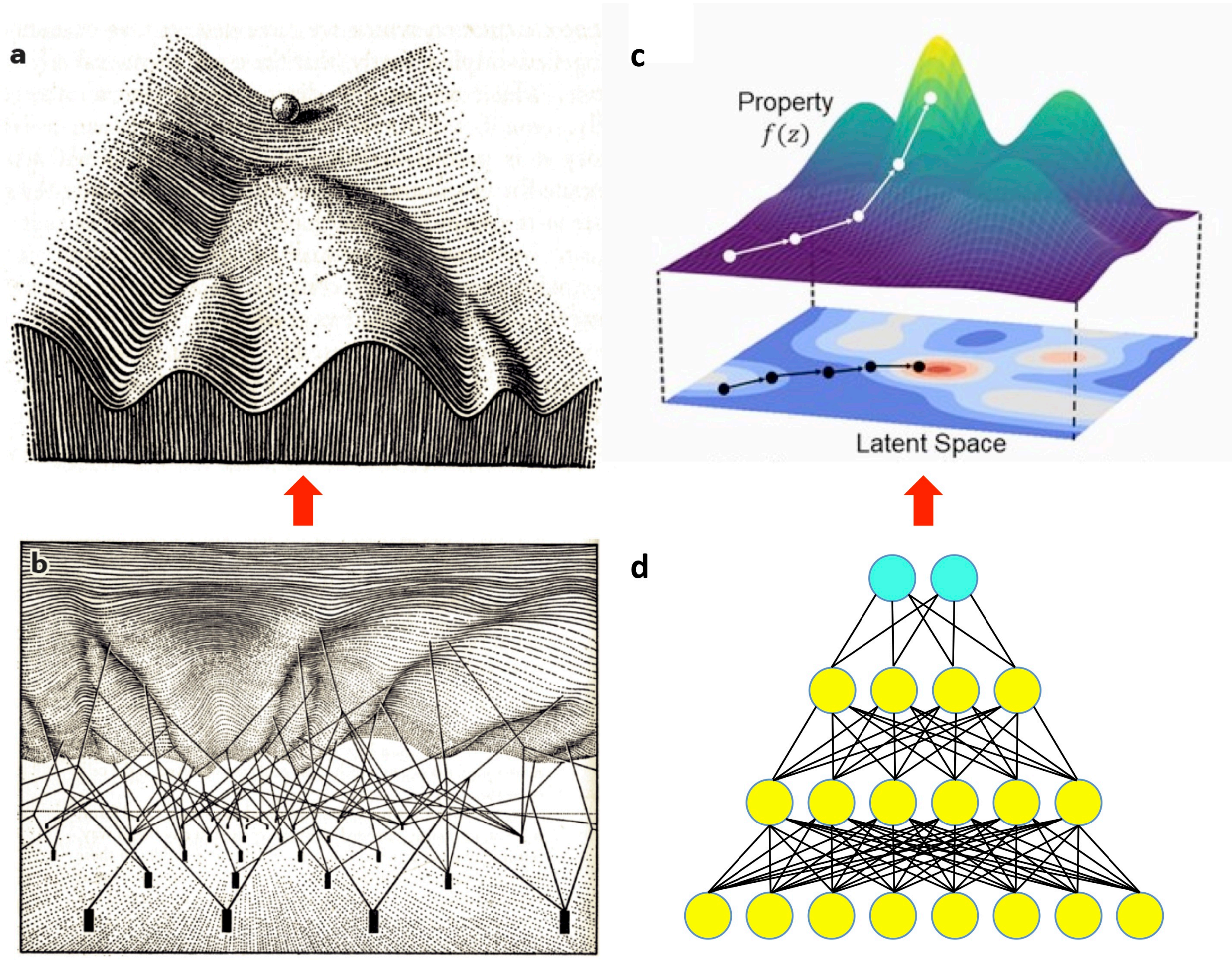

**Figure 3. Energy landscapes.** Panel (a) shows Waddington's epigenetic landscape, depicting the state of a cell through development as a ball rolling down a landscape, with the valleys representing eventual possible cell fates. One key aspect of this scheme is that it is probabilistic and statistical. The landscape is shaped by the collective actions of many genes (depicted in (b) as pegs and guy ropes pulling down on a canvas), but the actual outcome of any "run" for an individual cell will reflect the playing out of noisy processes over this landscape. Panel (c) represents a latent space and resultant energy landscape, generated from a compressed representation in a connectionist network (d). (Panels (a) and (b) reproduced, with permission, from [51]. Panel (c) reproduced, with permission, from [48].

Note that we are not talking here about traditional *fitness landscapes* used in evolutionary biology [45,52], though the two ideas are related. The energy landscapes created by generative models depict the states that the developing system will be channeled into by the constraint regimes of the model, whether defined in terms of cell types during differentiation, or eventual organismal phenotypes. The landscapes used in evolutionary biology depict a parameter at the ecological level: *fitness peaks* - i.e., how well-suited different phenotypes are to conditions [53].,The shape of the phenotype generation landscape reflects developmental biases – i.e., which phenotypes can be generated by current genotypes or reached by genetic change. The shape of the fitness landscape, in turn, reflects how well various phenotypes, including potentially novel ones, do in the current ecology. Together, these landscapes influence how evolution acts on the generative model itself - i.e., how it exerts selection over combinations of genetic variants.

**The encoder: evolution as a learning process**

Collectively, the latent variables in the genome can be thought of as constituting a connectionist network, with individual genes as nodes and regulatory interactions as edges (which can be activating or repressing), with weights determined by the relative affinities. An important question, then, is how these weights get set.

In the analogy to deep learning systems, evolution plays the role of learning or training. The idea that evolution is effectively a learning algorithm is not new, of course [34,54-57], but recent comparisons with the various algorithms of deep learning in neural networks are revealing some deep correspondences [58]. Though the learning mechanisms themselves may differ (typically back-propagation in ANNs, for example, versus variation followed by natural selection), the result is changes to the weights of the network [56].

It is important to emphasise how indirect this learning process is. “Selection” reflects the differences in reproductive success of organisms with different phenotypes. Those phenotypic differences may arise due to new mutations, which alter some weights in the network in some specific ways, or due to new *combinations* of genetic variants arising from sexual reproduction and recombination. Any differential reproductive success associated with the new phenotypes that emerge then results in an increased or decreased frequency of the variants in the relevant genomes in the next generation. The model encoded in the genome thus represents a historical record of information about the form of the organism and its fittedness to the environment.

This kind of evolutionary connectionism [59] naturally leads to a compressed encoding between model space and feature space (i.e., between genotype and phenotype). Compression is driven by the cost of adaptation in lineages and populations [60] and by “connection costs” in the network itself [61]. These connection costs naturally lead to modularity and, therefore, evolvability [61,62]. This emergent modularity is reinforced by environments where different factors vary independently over time, making separable adaptations optimal [30,61,63,64].

Crucially, this kind of indirect (generative) encoding, using compressed representations in the latent variable space of the genome, is more robust and more evolvable than direct representations of phenotype would be, as highlighted by Watson and Szathmary (2016) [56]:

"*By separating model space from feature space, learned models can be used to generate or identify novel examples with similar structural regularities, or (particularly relevant to evolution) to improve problem-solving or optimisation ability by changing the representation of solutions or reducing the dimensionality of a problem*"

There is an important point, however, in which biology and machine learning diverge. In machine learning, each new model starts with random weights and is trained *de novo* on large sets of data (e.g. images of horses), and develops a new compressed representation in the process. In biology, evolution has done the encoding job and each new individual inherits the compressed model (i.e., its genome). The encoding in biology is thus done across evolutionary time, in lineages, while the decoding is done by each new individual. Of course, what evolution has to act on – the only thing it can "see" – are the outcomes of these decodings, which will lead to some of the compressed generative models being favoured over others.

**Properties of the model**

A good test of this kind of conceptual analogy or model is whether it gives you anything more than what you put into it. In this case, the generative model analogy can account for and indeed predicts a number of additional genetic, developmental, and evolutionary phenomena. To explore these, we can begin by examining the important properties of generative models. These include:

1. Compression through a bottleneck layer.
2. Encoding in a latent variable space.
3. Abstract, indirect representations.
4. Intrinsic variability of outputs.
5. Robustness.
6. Evolvability.

These properties are tightly interrelated. Compression enforces the abstraction and encoding of latent variables. In machine learning, this prevents overfitting and yields the ability to generalize from the training data and create new instances of a given type. Crucially, this relies on a certain amount of randomness in the decoding process, which generates truly novel tokens. In machine learning models, such as variational autoencoders, the input data are represented as distributions from which individual values are sampled at random. The variability in outcome thus derives entirely from this representation of latent variables as distributions. In developing organisms, the randomness comes from the inherent noisiness of molecular processes – i.e., from the processes of decoding themselves. This noise is not just tolerated – it is an essential part of the system [7,10,29,65].

Collectively, these properties yield robustness and evolvability. The distributed nature of the genomic representations in latent variable space means that alteration of individual latent variables is often tolerated by the system. Indeed, the inherent noise in the system means the decoder always has to contend with some variability and still be able to robustly produce an outcome within the desired range. There is a clear analogy here to "denoising" autoencoders, which can take a noisy or staticky image and generate a clean one [27,28]. Similarly, "the attractor feature of an autoassociative [gene regulatory] network means that it can solve the problem of recovering a particular state (usually represented as a vector), when presented with an initial pattern that resembles one of the memory vectors stored in its weights" [66].

There is strong pressure, therefore, to not just encode developmental processes leading to adaptive outcomes, but to find architectures that do so robustly [67-69]. Paradoxically, this robustness leads to evolvability [30]. If no variation in parameters were ever tolerated, no change would be possible. In living organisms, many changes to the latent variables of the generative model (i.e., the DNA sequence) are well tolerated. This means genetic variation can accumulate in populations. Over time, however, these variants can have collective effects on the features of the output (i.e., the form or phenotype of the organism). In addition, the possibilities for phenotypic change are increased by the mixing and recombination of these variants during sexual reproduction [66,70]. This accumulated genetic variation is then the substrate for evolution and can be selected based on the relative fitness of the resultant phenotypes.

It might seem, however, that the distributed nature of the latent representations of the features of the encoded objects would militate against the possibilities for evolution to act on specific traits or aspects of the phenotype. If the energy landscape is shaped at every point by the actions of multiple genes, and if every gene pulls on the landscape at multiple points, it might seem impossible either to change anything at all or to change one thing without changing many others at the same time. Here, lessons from machine learning and neuroscience may also be enlightening.

**Emergent modularity and disentangled representations**

So far, we have been considering how a genome could encode the form of an organism, in a normative sense – a cat versus a dog, for example. But of course we also want to know how *variation* in the genome can lead to *variation* in phenotypes, within a given species. Conceiving of the genome as encoding a generative model may give us a more accurate and useful picture of the relationship between genotypes and phenotypes than alternative metaphors, such as a program or a blueprint.

While some single mutations may have large effects on various phenotypes, most phenotypic variation observed in natural populations reflects the combined effects of many genetic variants with individually tiny effects. That is, most traits are highly polygenic, or even omnigenic – at least potentially affected by variants in every gene expressed in a relevant tissue [71]. At the same time, most individual variants are pleiotropic – they may be statistically associated with variation in multiple phenotypic traits [72,73]. In addition, the effects of any individual variant, at a biological level, are often epistatic – that is, they may show a non-linear context-dependence on the presence of other genetic variants [73,74].

This architecture presents a puzzle. How could natural or artificial selection act on specific traits, without affecting all kinds of other traits at the same time? Work on generative machine learning models has shown that a distributed, non-linear encoding in latent variable space need not imply gridlock. Even if individual variables do not specifically or directly encode particular features, combinations of such variables may demonstrate what we call *emergent modularity*.

This feature is evident in many machine-learning models, especially ones where it has been selected for such that individual features in the output can be modified independently [75,76]. Different features may be represented in orthogonal low-dimensional subspaces or manifolds in latent variable space, thus making it possible to manipulate them independently [77]. For example, in some models designed to produce new images of human faces, it is possible to independently modify distinct features of the output, such as sex, age, facial hair, facial expression, accessories such as glasses, and so on [18, 78-81]. Critically, the probabilistic nature of latent spaces in many generative models enables the manipulated latent space to interpolate realistically between previously seen exemplars [82, 83]. Readers familiar with neuroscience may recognise the related ideas of disentangled representations, manifolds, and communication subspaces, the emergence of which is similarly favored under conditions where tasks vary independently [24,84-88].

The same of course is observed in animal and plant breeding. It is possible to select for an increase or decrease in specific quantitative traits by selective breeding, which enriches for combinations of genetic variants with shared effects on the selected trait, without altering the mean of all the other traits that the individual variants involved may also affect. To put this in machine-learning terms, the representations of different traits may be orthogonal to each other in latent variable space. It may thus be possible to push the output features along one dimension without altering the features on other dimensions (Figure 4). This kind of emergent modularity, with disentangled representations, may arise naturally under conditions where environmental factors themselves have a history of varying independently [54,55,61-64,89].

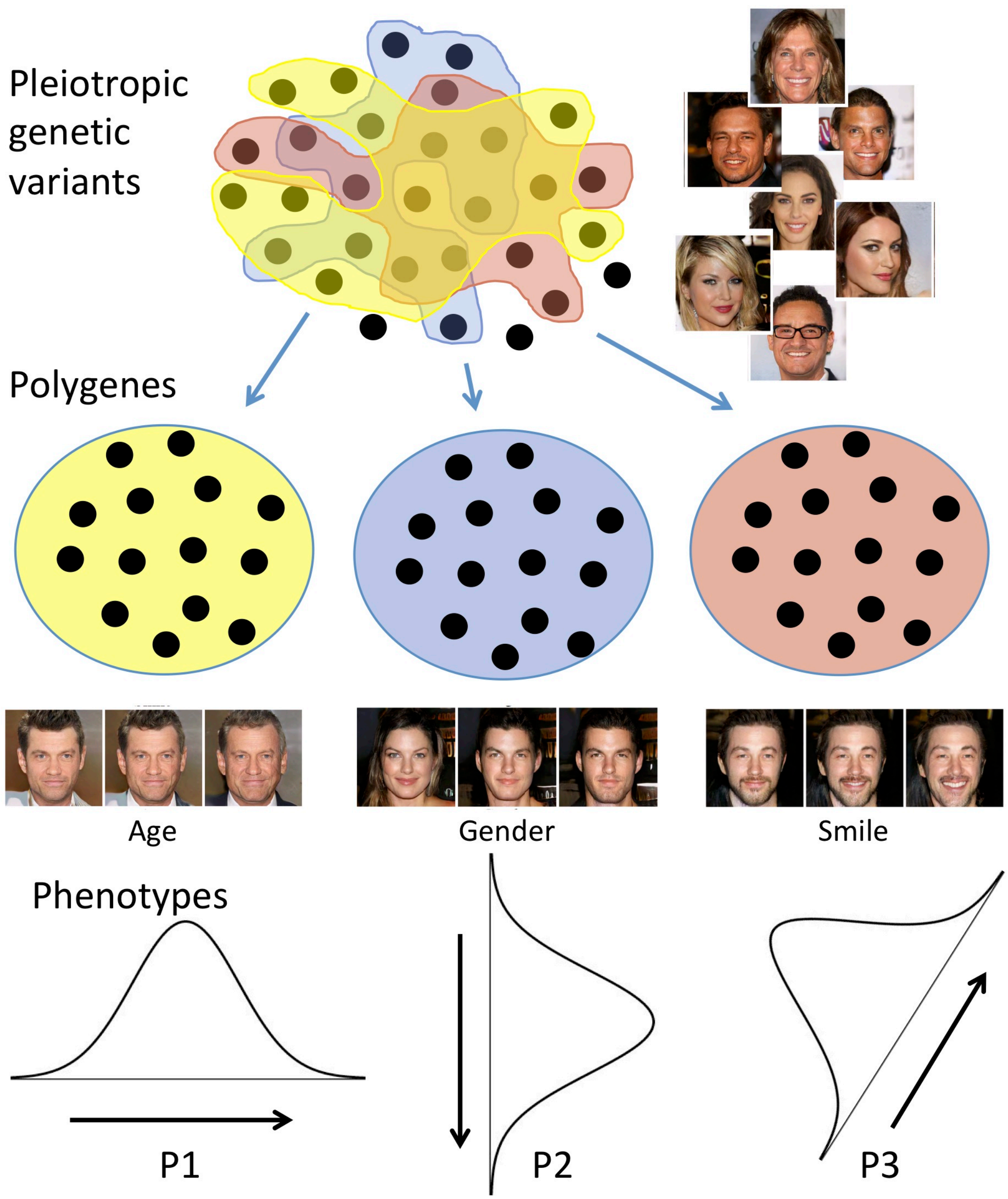

**Figure 4. Emergent modularity in genotype-phenotype relations.** Each genome carries millions of individual genetic variants, each of which may – in a tiny way – affect multiple different phenotypes (top row). Analogously images in the full-dimensional input or output space of a VAE carry a plethora of possible variations (top row). In turn, each phenotype is polygenic – i.e., affected by thousands of individual variants. However, these sets of variants (or "polygenes") may give rise to disentangled representations in latent variable space (middle row). In the analogy with VAEs for image generation, distinct features of encoded faces can be manipulated independently because they are encoded in orthogonal dimensions of latent variable space (from [81]). In living organisms, this kind of orthogonal encoding means that different phenotypes (P1-P3) may be selected independently (bottom row).

### Alternate Attractor States

The generative model framework can readily accommodate the observation that ***alternate phenotypes*** can be encoded from the identical genome in a flexible, environmentally responsive fashion. Examples of this kind of phenotypic plasticity include sex determination in response to termperature in some amphibians and reptiles [90], as well as the production of workers, soldiers, queens, and other specialised types in many species of ants, bees, and other social insects [91]. In these species, the generative model effectively encodes, in a multiplexed fashion, multiple possible anatomical forms with characteristic physiological and behavioural profiles, with the form which emerges depending on an interplay with environmental factors [91].

Finally, complex dynamical systems may also display unanticipated, qualitatively novel, emergent phenotypes. The non-linearities inherent in generating energy landscapes that constrain developmental processes can – under altered circumstances – sometimes generate novel, but typically latent attractor states, pulling the system into points in state space that are far from those normally occupied. Cancer is an obvious (maladaptive) example, at the molecular level, but there could also exist novel attractors at the level of organismal form. The states that may emerge under these kinds of circumstances are a property of the architecture of latent variable space, but one that has not been trained or directly selected for. Such qualitatively novel (and often maladaptive) phenotypes are much harder to explain in blueprint or program conceptions.

### Formalisability

A key advantage to the generative model conception is that it can readily be formalised. Waddington's perspective remained metaphorical while developmental biology was limited to reductive approaches aimed at manipulating single genes at a time [92]. But this old idea is now being revived and formalised in many areas of developmental biology, as researchers finally have the tools to investigate whole systems [32,47,93]. These include the ability to monitor expression of all genes in the genome in single cells, as well as the computational tools and concepts to model the resultant data [34,47,66]. The success of these approaches reinforces the view that Waddington's landscape is not just a useful way of thinking about how the genome directs the processes of development – it is the right way.

The regulatory interactions underlying a gene regulatory network can be formally modelled with sets of ordinary or partial differential equations [94]. Setting these by hand becomes cumbersome or effectively impossible after a certain (fairly low) level of complexity. Clustering and machine learning methods have also been used to implement dimensionality reduction from large amounts of single-cell gene expression data, in order to filter the noise and derive interpretable models of the underlying gene regulatory network and cell fate transitions for subsequent functional analysis in silico [34,94-96]. Recent work has also taken advantage of deep learning methods [97-99], including variational autoencoders [96-98,100-102] for similar purposes.

Maizels et al. [100] note that “deep generative modelling is particularly well suited to learning hidden variables that capture the complex distributions within high dimensional data.” We propose a deeper correspondence: that deep learning connectionist methods that produce generative models are not just a useful tool for analysing or simulating gene regulatory networks but reflect more fundamentally the principles by which such networks come to be in the first place. The genome instantiates just such a compressed, generative model, with latent variables learned through evolution [56] and decompressed through the processes of development to generate new individuals.

This perspective may also have direct applicability in the field of Artificial Life (ALife). Various types of encoding schemes are currently used in this field, to set a relationship between an artificial “genome” and the form of an artificial organism, whether that is realized in physical form or in silico. The vast majority of ALife encoding schemes, whether simple direct encodings or complex high-non-linear or developmental encodings, tend to be manually predefined by the experimenter before machine learning tools are used to find high-performing genomes for those specified developmental rules [103-107]. The perspective presented here, of framing an ALife genome as analogous to a latent space vector in a deep neural network, helps to enable a growing body of work that uses the machine learning methods, including variational autoencoders, that train deep neural networks to find high performing encoders and decoders for this genotype-to-phenotype mapping [108-117].

One important point to bear in mind in any such attempt to evolve genomic encodings of artifcial “life-forms” *in silico*, is that, for natural life-forms, the world itself provides crucial context. The genome does not encode the physics of molecules and cells and tissues or their self-organising tendencies – it merely constrains them. Any encodings in a virtual world will therefore require a context of virtual physics in which to operate.

## Conclusion

The history of biology highlights the danger of naively reaching for the latest technological fad as a metaphor for some aspect of biology [118]. However, thinking of the genome as instantiating a generative model of the form of the organism seems fairly well supported by the analyses presented above. Of course, the metaphor has limits, notably in the way the model is acquired and in the architecture of the network that encodes it.

We should emphasise that we are not arguing that the genome itself, or reproducing, evolving biological systems more broadly are a type of VAE. Rather, we argue that both the natural and artificial systems have some abstract principles in common. Indeed, one could reverse the analogy and argue that human designers have independently hit upon learning algorithms and architectures that rely on forming a compressed representation in a space of latent variables in a way that seems to be similar, in abstract terms, to how the genome represents the form of the organism. Furthermore, while we picked the VAE architecture in order to make the analogy concrete, one could draw similar comparisons with other machine learning systems, such as Generative Adversarial Networks [119], or to other more developmentally inspired decoders, such as such as Neural Cellular Automata [120,121]. By arguing that both VAEs and biological organisms are instances of generative models, we aim to highlight the correspondences between the abstract principles at play, rather than the detailed instantiations.

The important question is whether the analogy, as we have presented it is conceptually and heuristically useful [1,118]. We argue that it is an improvement over the concepts of the genome encoding a “blueprint”, “program”, “recipe”, or “resource”. The first two imply an overly linear, isomorphic, direct, and deterministic relationship between aspects of genotype and phenotype or the developmental processes that produce phenotypes. The third is more organic, but does not offer meaningful mechanistic insights, while the fourth is simply rather vague. The “generative model” concept avoids those connotations and seems to give a more apt conception of the genotype-phenotype relationship and the nature of information encoding in the genome. It may indeed be the foundation of a theory, rather than just a metaphor or conceptual analogy.

In particular, the notion of the genome encoding a generative model for the development of a new individual organism seems to capture very well the job description of the genetic material. Of course, development requires many additional resources – most notably, a fertilised egg with the appropriate cytoplasmic endowments, but also a permissive environment more widely. Indeed, one could argue, along lines proposed by Developmental Systems Theory [8], that it is more appropriate to think of the genome and the fertilised egg *together* as the generative model, with the genome *per se* just embodying the compressed representation. In either framing, the important point is that the genome plays a distinctive role in this process and it seems perfectly apt to describe it as a store of information about organismal form (and its fittedness to the environment), and, at the same time, a store of information about the machinery necessary for the execution of the developmental process. Indeed, the generative model concept may help us move from a view where the genome somehow actively *drives* development to one where the latent variables of the model collectively constrain the self-organising processes of development.

The encoder-model-decoder scheme also matches well the processes of evolution, compression in each generation to a single cell, and decompression of this information through development. Moreover, the known properties of artificial autoencoder systems, including the generation of compressed, orthogonal representations can account for a number of observed properties of genotype-phenotype relations, including the evolvability of distinct traits. However, the mark of a useful analogy is that it also highlights interesting *differences* between the analogised systems or points that need fleshing out, which can act as a spur to further thought (see Outstanding Questions).

However, what the analogy adds in precision, it lacks in familiarity. If the goal is to explain these concepts to the general public, appealing to an unfamiliar term like a "generative model" is unlikely to be immediately helpful. To communicate this idea effectively will likely require a good deal of explication. But perhaps we should not be surprised that one of the deepest mysteries of life – how the genome encodes the form of an organism – cannot, in fact, be captured and conveyed by a simple, familiar word or phrase or concept.

**Acknowledgments**

We thank Henry Potter, Shawn Beaulieu and Nate Gaylinn for very helpful discussions and comments on the manuscript. This material is based upon work supported by the National Science Foundation under Grant No. 2239691.

**Declaration of Interests:** The authors declare no conflicting interests.